\documentclass{appolb}
\usepackage{epsfig}
\usepackage{slashed}
\usepackage{mathrsfs}
\usepackage{graphicx}
\usepackage[usenames]{color}
\usepackage[colorlinks=true,linkcolor=blue,citecolor=blue,urlcolor=black]{hyperref}
\usepackage{dcolumn}
\usepackage{subfigure}
\usepackage{xspace}
\usepackage{array}
\usepackage{enumerate}

\newcommand{\ua}{\ensuremath{U(1)_A}}
\newcommand{\vev}[1]{\ensuremath{\left\langle #1 \right\rangle}}

\begin{document}
\pagestyle{plain}
\title{Three-flavor chiral phase transition and axial symmetry
  breaking with the functional renormalization group\thanks{Presented
    at the XXXI Max Born Symposium and HIC for FAIR Workshop {\it
      ``Three days of critical behaviour in hot and dense QCD''},
    Wroclaw, Poland, June 14-16, 2013}} \author{Bernd-Jochen Schaefer
  \address{Institut f\"{u}r Theoretische Physik,
    Justus-Liebig-Universit\"at Giessen,
  Heinrich-Buff-Ring 16, D-35392 Giessen, Germany}
\and
Mario Mitter
\address{ Institut f\"ur Theoretische Physik, Universit\"at Heidelberg,\\
   Philosophenweg 16, D-69120 Heidelberg, Germany}
  \address{\centering  Institut f\"ur Theoretische Physik, Goethe-Universit\"at Frankfurt, 
  Max-von-Laue-Stra{\ss}e 1 , D-60438 Frankfurt/Main, Germany}
}
\maketitle
\begin{abstract}
  The interplay of mesonic fluctuations with an axial $\ua$-symmetry
  breaking and resulting effects on the location of a possibly
  existing critical endpoint in the QCD phase diagram are investigated
  in a framework of the functional renormalization group within a $N_f
  = 2+1$ flavor quark-meson model truncation. The axial $\ua$-symmetry
  breaking is imposed by a mesonic Kobayashi-Maskawa-'t~Hooft
  determinant. The quark mass sensitivity of the chiral phase
  transition with and without the $\ua$-symmetry breaking is studied.
\end{abstract}

\bibliographystyle{./bibstyle}
\graphicspath{
{./}
 }
  
\newpage

\section{Motivation}
It is well-established knowledge that Quantum Chromodynamics (QCD)
experiences a rapid crossover from a hadronic phase with broken chiral
symmetry to a deconfined and chirally-symmetric quark-gluon plasma at
high temperature and moderate baryon densities. In the chiral limit,
i.e., for $N_f$ massless quark flavors the QCD Lagrangian is invariant
under $U(1)_V\otimes U(1)_A\otimes SU(N_f)_L\otimes SU(N_f)_R$
transformations where for our discussion unimportant discrete
subgroups have been ignored. The vector subgroup $U(1)_V$ corresponds
to quark-number conservation and seems to be less important at the
chiral transition.  However, it is uncertain what happens to the
anomalously broken $\ua$-symmetry at the transition.  If the
$\ua$-symmetry is broken the relevant symmetry is reduced to the
chiral $SU(N_f)_L \otimes SU(N_f)_R$ which is for two massless flavors
isomorphic to the $O(4)$ symmetry. For a continuous phase transition
this would lead to a three-dimensional $O(4)$ universality
class. However, the symmetry-breaking pattern changes significantly
with a restored $\ua$-symmetry leading to an essential impact on the
nature of the chiral phase transition.

Recent experimental observations found a drop of at least $200$ MeV in
the anomalously large mass of the $\eta'$-meson close to the chiral
crossover which might signal an effective restoration of the
$\ua$-symmetry \cite{Csorgo2010,Vertesi:2009wf}.  On the theoretical side, the
situation is more controversial: A number of recent QCD lattice
simulations found a substantial suppression of $\ua$-anomaly related
effects around the crossover \cite{Bazavov:2012qja, Cossu:2012gm,
  Cossu:2013uua} but there are also very recent results indicating a
$\ua$-symmetry breaking in terms of chiral susceptibilities above the
crossover \cite{Buchoff:2013nra}.  Furthermore, some analytical
studies show an effective $\ua$-symmetry restoration at the transition
in the chiral limit for three quark flavors but not for two flavors
\cite{Cohen:1996ng, Lee:1996zy, Birse:1996dx, Dunne:2010gd}, see also
\cite{Meggiolaro:2013swa}.  Finally, the order and universality
class of the chiral transition in the chiral limit for two quark
flavors are not fully settled and depend on the behavior of
$\ua$-symmetry breaking operators at the transition
\cite{Pisarski1984a, Grahl:2013pba, Pelissetto:2013hqa}.

In the following the role of the $\ua$-symmetry breaking at
non-vanish\-ing temperatures and quark chemical potentials is
addressed.  Fluctuations are taken into account by solving functional
renormalization group (FRG) equations in a three flavor quark-meson
model truncation. The $\ua$-symmetry breaking is implemented
effectively by a mesonic Kobayashi-Maskawa-'t~Hooft determinant in the
truncation \cite{Kobayashi:1970ji, 'tHooft:1976fv}.  Particular focus
will be put on the interplay of mesonic fluctuations and
$\ua$-symmetry breaking. Consequences for the location of a
possibly existing critical endpoint in the QCD phase diagram are
discussed and the quark mass sensitivity of the three flavor chiral
transition with and without $\ua$-symmetry breaking is explored.

\section{Flow equations for the three-flavor quark-meson model}

Low-energy QCD with $N_f=3$ quark flavors can be described with an
effective chiral quark-meson model which captures the degrees of
freedom of strongly-interacting matter relevant for the chiral phase
transition.  The Euclidean Lagrangian \cite{Schaefer:2008hk}
\begin{eqnarray}\label{eq:lagrangian}
  \mathcal{L} &= \bar q \left[\slashed{\partial}+ \mu \gamma_4 +
    h\left(\sigma_b+ 
      i\gamma_5\pi_b\right) T^b\right] q +
  \mathrm{Tr}\left[ \partial_\mu \Sigma^\dag\partial_\mu \Sigma\right]
  + \tilde U(\rho_1, \tilde\rho_2)\ ,
\end{eqnarray}
includes two degenerate light and one strange quarks $q$ which are
coupled through a flavor-independent Yukawa coupling $h$ to scalar,
$\sigma_b$, and pseudoscalar mesons $\pi_b$.  Additionally, we have
introduced a flavor symmetric quark chemical potential $\mu$.  The
purely mesonic theory is parameterized with fields in matrix form
$\Sigma=(\sigma_b+i\pi_b)T^b$ where $T^b$ are the generators of the
flavor $U(3)$ group. The potential $\tilde U(\rho_1,\tilde\rho_2)$
parameterizes the interactions of all eighteen mesons in terms of the
chiral invariants
\begin{eqnarray}\label{eq:chiral_invariants}
 \rho_1= \mathrm{Tr}\left[\Sigma^\dag\Sigma\right]\ ,\quad \tilde\rho_2= \mathrm{Tr}\left[\left(\Sigma^\dag\Sigma\right)^2\right]-\frac{\rho_1^2}{3}\ .
\end{eqnarray}

Non-vanishing quark masses can be implemented in the quark-meson model
with explicit symmetry breaking terms that are linear in the uncharged
scalar mesons $\sigma_0$, $\sigma_3$ and $\sigma_8$. Since the
$\sigma_3$ term breaks the $SU(2)$-isospin symmetry explicitly, the
$(2+1)$-flavor version of the quark-meson model is obtained by
ignoring the $\sigma_3$ and consider only the $\sigma_0$ and
$\sigma_8$ breaking. For explicit calculations it is convenient to
change the singlet-octet basis and perform a unitary rotation to the
non-strange, $\sigma_x$, and strange, $\sigma_y$, scalar fields.

The Lagrangian defined in Eq.~(\ref{eq:lagrangian}) respects the full
chiral symmetry $U(1)_V\times SU(3)_L\times SU(3)_L\times \ua$.
The anomalous breaking of the $\ua$-symmetry can be taken into
account by adding a Kobayashi-Maskawa-'t~Hooft determinant
\cite{Kobayashi:1970ji, 'tHooft:1976fv}
\begin{eqnarray}\label{eq:determinant}
 \xi &= & \det\Sigma + \det\Sigma^\dag\ ,
\end{eqnarray}
which is cubic in the (pseudo)scalar meson fields, to the Lagrangian
Eq.~(\ref{eq:lagrangian}).  This interaction is invariant under $U(1)_V\otimes
SU(3)_L\otimes SU(3)_L$ and breaks the $U(1)_A$-symmetry in
Eq.~(\ref{eq:lagrangian}).  With this term the proper mass splitting
between $\eta$- and $\eta'$-meson as well as the pion mass can be
reproduced but other implementations of the $\ua$-symmetry breaking
are in general possible.  The breaking term in
Eq.~(\ref{eq:determinant}) scales with the meson fields to the power
$N_f$ which leads to important $N_f$-dependent effects for the chiral
phase transition.  For three flavors, the mass splitting effects,
caused by the cubic determinant, disappear in a symmetric phase of
vanishing expectation values for all mesonic fields.

In total the chirally invariant meson potential of the $(2+1)$-flavor
quark-meson model $\tilde U$ in Eq.~(\ref{eq:lagrangian}) is replaced
by
\begin{eqnarray}\label{eq:mes_pot}
 \tilde U(\rho_1,\tilde\rho_2)&\to & \tilde U(\rho_1,\tilde\rho_2)-
 c\,\xi - c_x \sigma_x - c_y \sigma_y\ , 
\end{eqnarray}
where two explicit chiral and one $\ua$-symmetry breaking terms have
been added.

For a non-perturbative renormalization group analysis of the chiral
phase transition we employ the effective average action approach by
Wetterich \cite{Wetterich:1992yh}. We truncate the effective action in
form of a quark-meson model in a leading order derivative
expansion. Recently, the relation of these truncations to full QCD has
been affirmed by using dynamical hadronization. This approach
introduces RG scale-dependent mesonic degrees of freedom that 
eliminate four-fermion interactions generated by gluon exchanges
\cite{Gies:2001nw, Gies:2002hq, Floerchinger:2009uf,
  Pawlowski:2010ht}, see also \cite{Braun:2011pp}.
Extending the formalism to finite temperature $T$, the flow equation
for the RG-scale $k$ dependent and symmetrical potential $\tilde{U}_k$
with an optimized three-dimensional regulator \cite{Litim:2001up}
reads
\begin{eqnarray}\label{eq:flow}
    k\frac{\partial\tilde U_k}{\partial_k } &=& \frac{k^5}{12\pi^2} \Bigg[ 
	  \sum\limits_{b=1}^{2N_f^2} \frac{ 1}{E_b}\coth\left(\frac{ E_b}{2T}\right)  
	  \\
    && \phantom{\frac{k^5}{12\pi^2} \Bigg[}	  - 6 \sum\limits_{f=1}^{N_f} \frac{1}{E_f} 
      \Biggl\{ \tanh\left( \frac{E_f + \mu}{2T}\right) + \tanh\left( \frac{E_f - \mu}{2T}\right) 
	  \Biggr\} \Bigg]\ .\nonumber
\end{eqnarray}
The fermionic ($f$) and bosonic ($b$) quasi-particle energies have the
typical form $E_{i}=\sqrt{k^2+m_{i}^2}$. Explicit expressions for the
corresponding meson masses $m^2_b$, calculated with the potential
Eq.~(\ref{eq:mes_pot}), can be found in \cite{Mitter:2013fxa}.  In the
nonstrange-strange ($x-y$) basis the quark masses simplify to $m_x = h
\sigma_x/2$ and $m_y = h \sigma_y/\sqrt{2}$.

The solution of this flow equation describes the scale evolution of
the effective potential starting from an initial potential at some
high ultraviolet scale $\Lambda$ towards the full quantum effective
potential in the infrared $k\rightarrow 0$ \cite{Schaefer:2006sr}.
Evaluating the evolved potential at the minimum yields the grand
potential as a function of temperature $T$ and quark chemical
potential $\mu$ which includes all quantum and thermal
fluctuations. We adjust the initial potential $\tilde{U}_{k=\Lambda}$
at the UV scale $\Lambda$ to reproduce known experimental observables
in the infrared such as the pion mass or decay constants in the vacuum
\cite{Mitter:2013fxa}.  Concerning our implementation of
$\ua$-symmetry breaking we consider two different scenarios: one with
a constant, i.e. temperature and RG scale independent coupling for the
Kobayashi-Maskawa-'t~Hooft determinant and another without the
determinant.  In a future work we will improve this truncation by
considering a running version of this coupling \cite{mmanomalie} and
in \cite{Pawlowski:1996ch} the scale-dependency of the $\ua$-anomaly
induced determinant has already been analyzed in an RG context with
QCD degrees of freedom in the vacuum.

For comparison, we also present results obtained with a standard
mean-field approximation (MFA) where the mesonic fluctuations are
completely neglected and a divergent vacuum contribution from the
quark loop to the grand potential has been dropped. On the other hand,
this divergent vacuum contribution is included in the flow
equation~(\ref{eq:flow}). Hence, its influence on the phase transition
can be investigated by solving the flow equation without mesonic
fluctuations which is also called extended mean-field approximation.

\section{Chiral transition and axial anomaly}

\begin{figure*}[t!]
  \centering \subfigure[with $\ua$-symmetry
  breaking]{\includegraphics[width=6.2cm]{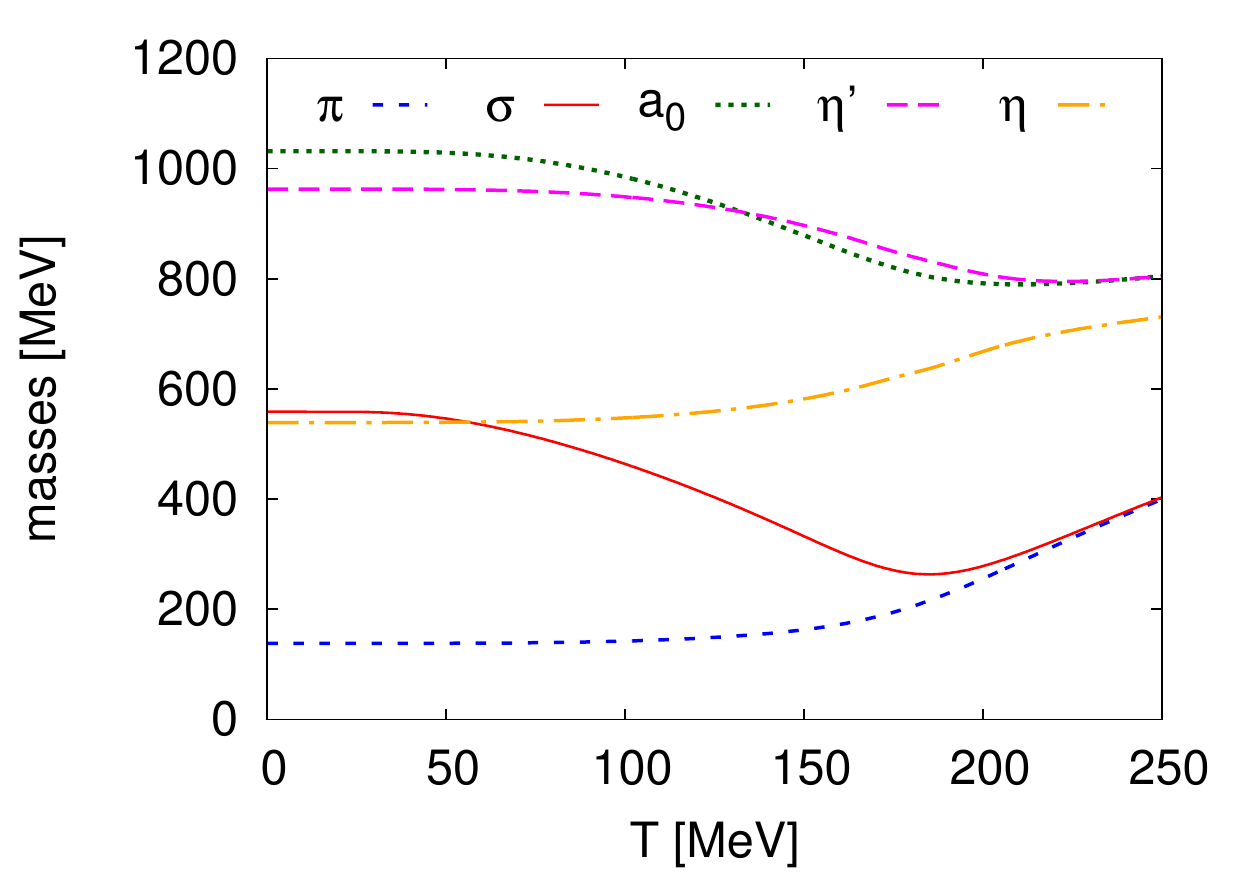}}
  \subfigure[without $\ua$-symmetry
  breaking]{\includegraphics[width=6.2cm]{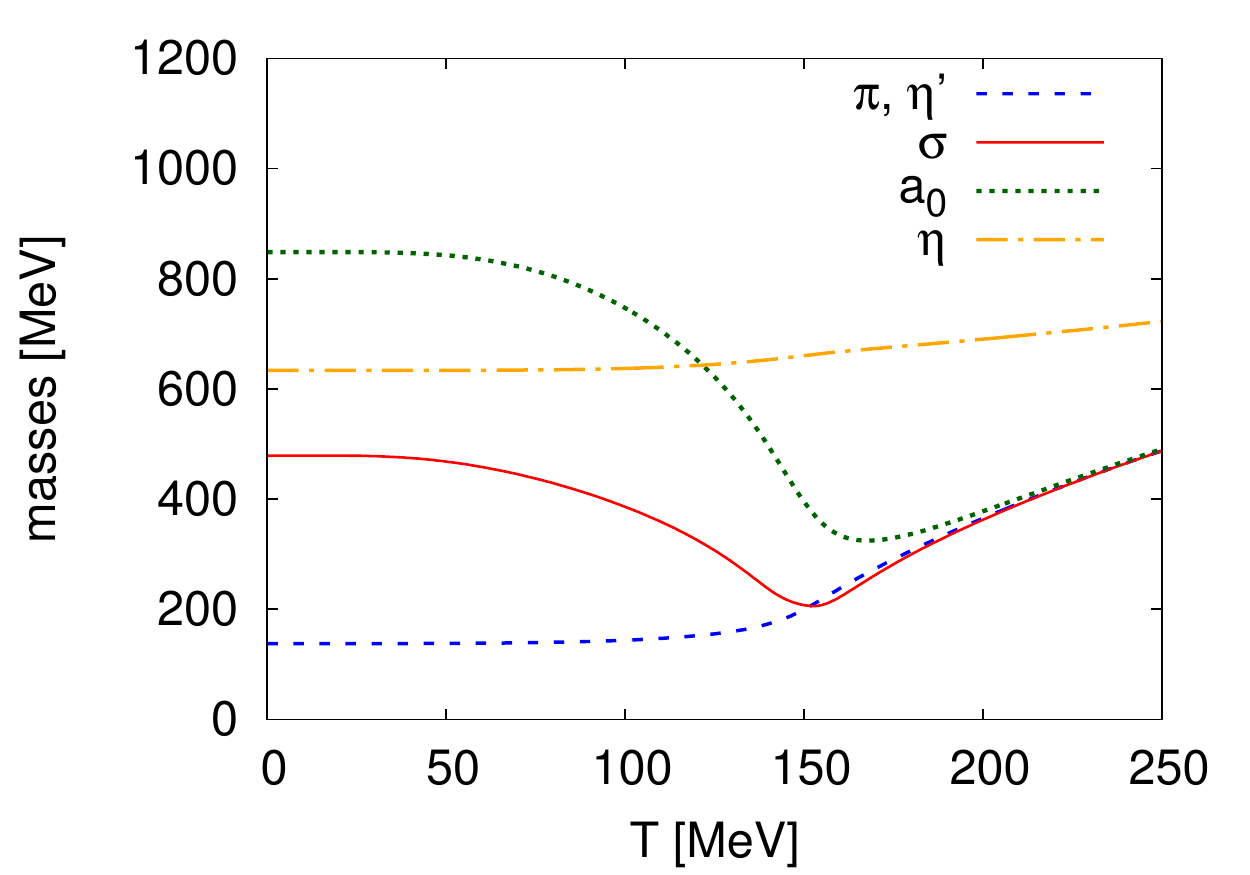}}
  \caption{\label{fig:frg_mf_meson_masses_one_phys_mu0} Scalar and
    pseudoscalar meson masses for $\mu=0$.}
\end{figure*}

We start with a discussion of the three-flavor chiral phase transition
at finite temperature and flavor symmetric chemical potential $\mu$
and investigate the interplay of fluctuations with the anomalous
$\ua$-symmetry breaking. The finite-temperature behavior of the
(pseudo)scalar non-strange and $\eta$-, $\eta'$-meson screening masses
obtained with the FRG are shown in
Fig.~(\ref{fig:frg_mf_meson_masses_one_phys_mu0}) for $\mu=0$ with
(left panel) and without (right panel) $U(1)_A$-symmetry breaking.
Without breaking the $\eta'$-meson degenerates with the pion mass and
the two sets of light chiral partners $(\sigma, \vec \pi)$ and $(\vec
a_0,\eta')$ merge in the chirally symmetric high-temperature phase.
On the other hand, for a constant temperature- and scale-independent
$\ua$-symmetry breaking a mass gap between the two sets of the chiral
partners is obtained and the $\eta'$-meson mass drops about $200$ MeV
near the chiral crossover in agreement with experimental observations
\cite{Csorgo2010,Vertesi:2009wf}.  Since the anomalous contribution to the
$\eta'$- and $\eta$-meson masses is proportional to the condensates
this behavior is a consequence of the melting of the light condensate
$\langle\sigma_x\rangle$ at the chiral crossover. Furthermore, the
melting of the light condensate entails that the $\eta$-meson is
dominantly strange and the $\eta'$-meson dominantly non-strange above
the crossover temperature \cite{Schaefer:2008hk}. The remaining difference between
$\eta'$-meson and pions is then directly proportional to the strange
condensate $\langle\sigma_y\rangle$ which decreases much slower.

\begin{figure*}[t!]
  \centering 
  \subfigure[with $\ua$-symmetry breaking]{\includegraphics[width=6.2cm]{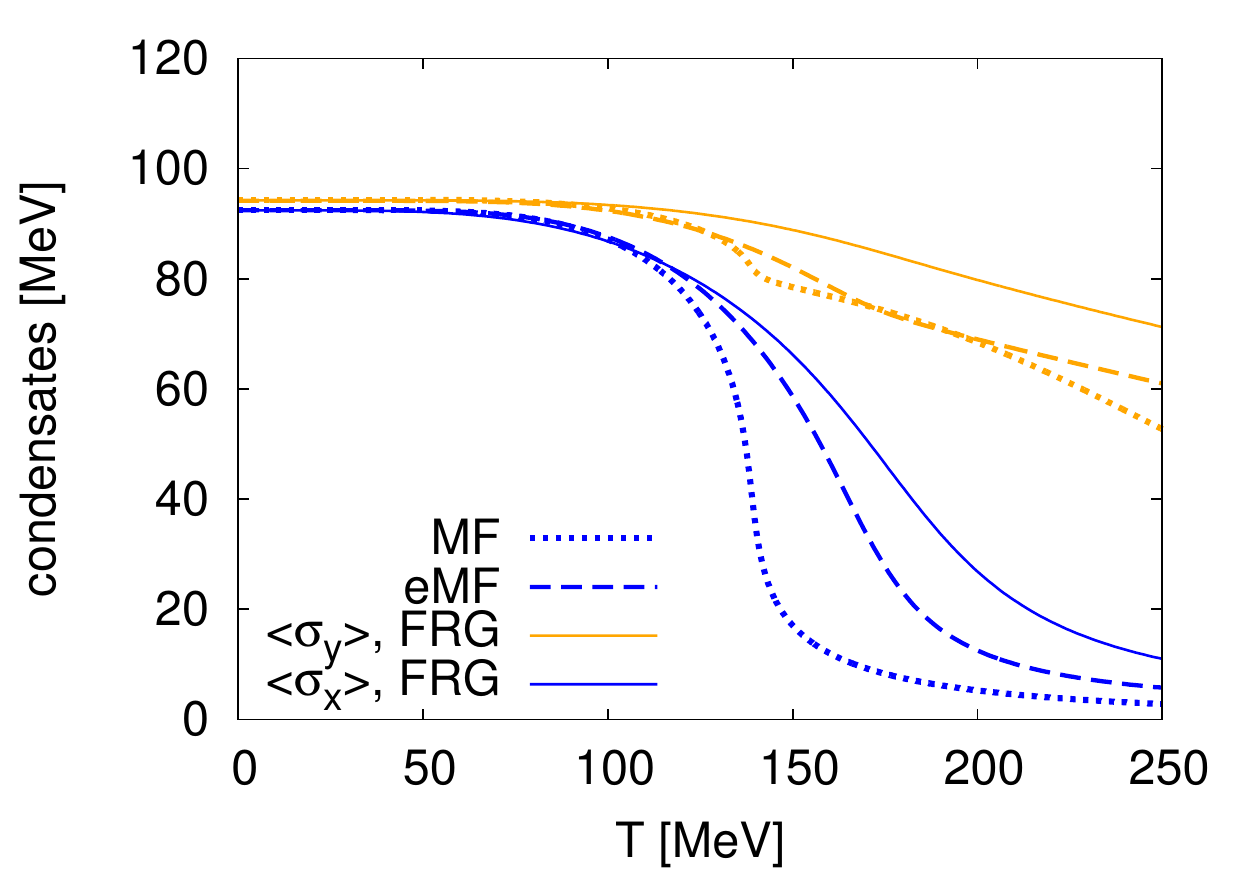}} 
  \hfill
  \subfigure[without $\ua$-symmetry breaking]{\includegraphics[width=6.2cm]{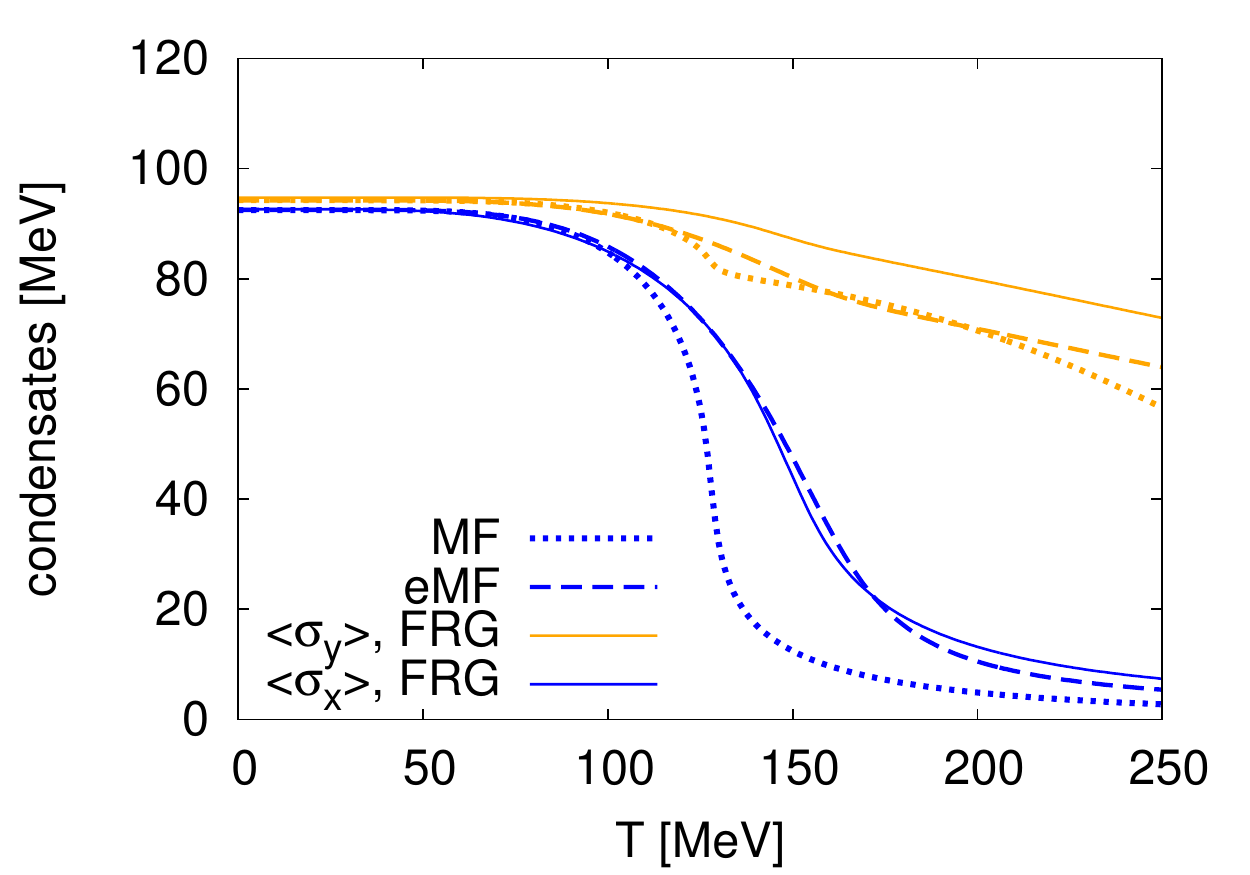}} 
  \caption{\label{fig:frg_condensates_phys_mu0} Non-strange
    $\vev{\sigma_x}$ and strange $\vev{\sigma_y}$ chiral condensates
    for $\mu=0$.}
\end{figure*}

The melting of both condensates as a function of temperature is
demonstrated for $\mu=0$ in Fig.~(\ref{fig:frg_condensates_phys_mu0})
again with (left panel) and without $\ua$-symmetry breaking (right
panel).  In addition to the FRG results (solid lines) we also show
results obtained with a standard MFA (dotted lines) and extended MFA
(dashed lines). In analogy to similar two flavor investigations
\cite{Schaefer:2006ds} we find generally that fluctuations wash out
the transition and the condensates decrease faster in the standard
MFA. However, this behavior cannot be solely attributed to mesonic
fluctuations. Comparing the FRG non-strange condensate
$\langle\sigma_x\rangle$ with the one obtained in the renormalized
model (eMFA) the influence of the $\ua$-symmetry breaking term becomes
visible. Without the breaking term both non-strange condensates agree
well whereas with a $\ua$-breaking term the melting of
$\langle\sigma_x\rangle$ is further softened if mesonic fluctuations
are added.  In other words, the chiral transition is considerably affected by
the Kobayashi-Maskawa-'t~Hooft term only if mesonic fluctuations are
taken into account whereas the mean-field investigations show only a
weak dependence on the $\ua$-symmetry breaking.

\begin{figure*}[t!]
  \centering
  \hfill
  \subfigure{\includegraphics[width=6.2cm]{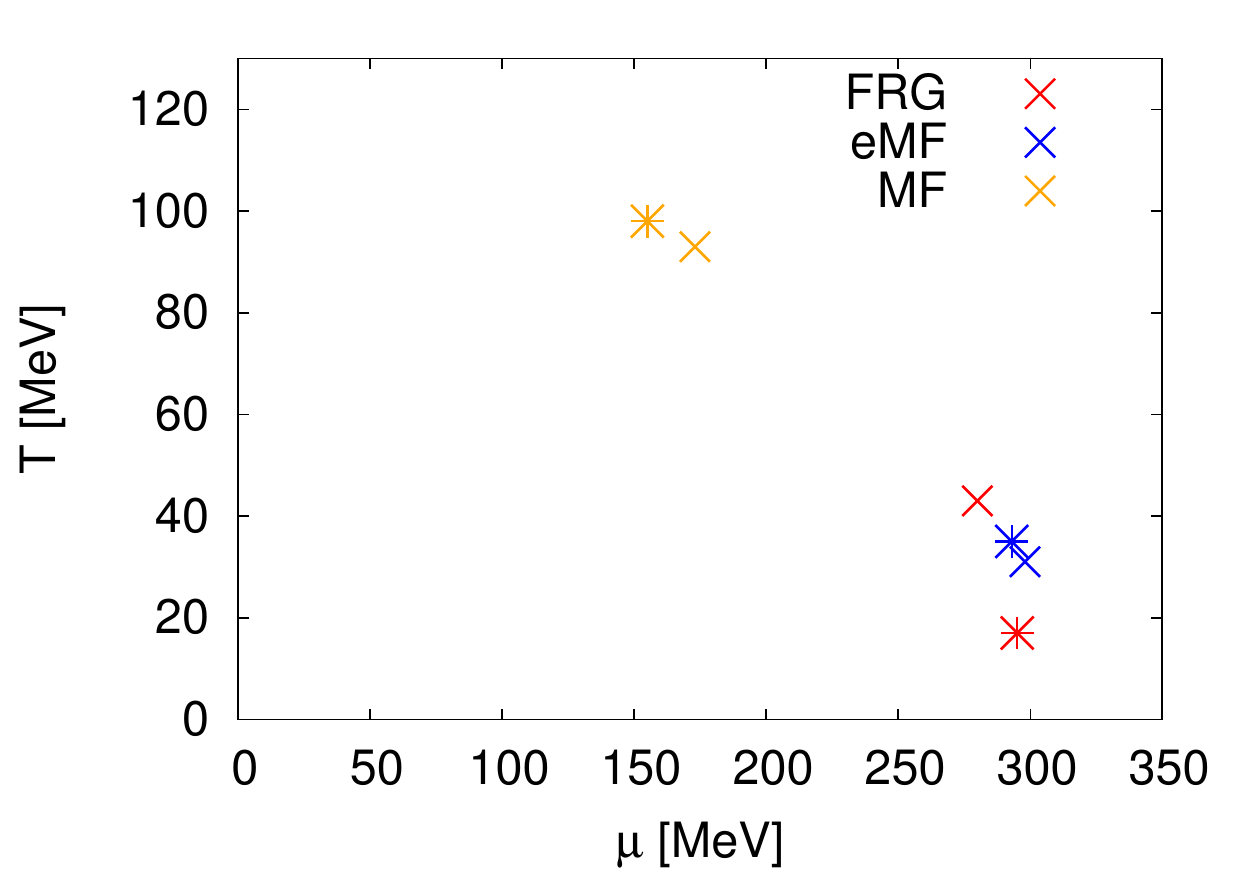}}
  \subfigure{\includegraphics[width=6.2cm]{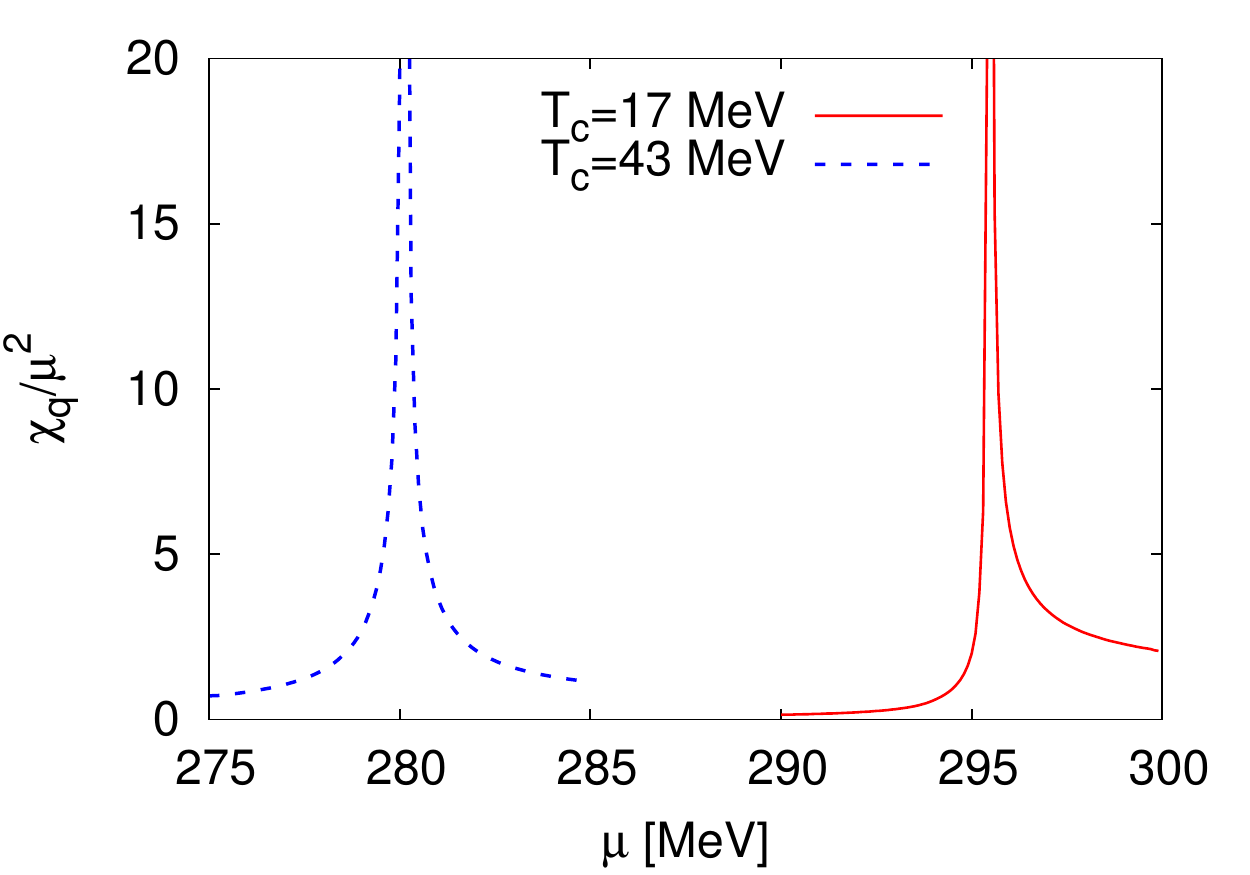}}
  \caption{\label{fig:frg_crit_exp_CEP} Charts of critical endpoints.
    Left panel: stars $\ua$-symmetry breaking included, crosses: no
    $\ua$-symmetry breaking term. Right panel: FRG quark number
    susceptibilities, dashed lines without and solid lines with
    $\ua$-symmetry breaking term. }
\end{figure*}

At non-vanishing chemical potential the impact of mesonic fluctuations
including a $\ua$-symmetry breaking has similar consequences. For this
purpose we investigate the existence and location of the critical
endpoint (CEP) in the $(T,\mu)$-plane obtained with the FRG and
mean-field approximations with and without the
Kobayashi-Maskawa-'t~Hooft determinant. The results are collected in
the left panel of Fig.~(\ref{fig:frg_crit_exp_CEP}). An endpoint
labeled with a star is the corresponding result including the
determinant and a cross denotes the results without a $\ua$-symmetry
breaking term. Since the location of the endpoints also depends
considerably on the chosen value of the $\sigma$-meson mass
\cite{Schaefer:2008hk} we have fixed the value to $m_\sigma =480$ MeV
in all calculations. At the critical point the quark-number
susceptibiliy diverges because the chiral transition is of
second-order. This is shown in the right panel of
Fig.~(\ref{fig:frg_crit_exp_CEP}).

Similar to the $\mu=0$ results, we see that the inclusion of the
fermionic vacuum term (eMF) makes the system less critical and
therefore the location of the CEP is pushed towards larger chemical
potentials and smaller temperatures as compared to the standard
mean-field approximation. In both mean-field approximations the
influence of $U(1)_A$-symmetry breaking is rather weak and we find
that the CEP without a breaking term is always at slightly smaller
temperatures and larger chemical potentials. Adding mesonic
fluctuations with the FRG leads to a qualitative change. With a
$U(1)_A$-symmetry breaking, the mesonic fluctuations push the CEP to
even larger chemical potentials and smaller temperatures in contrast
to the mean-field approximations. This behavior is in agreement with
observations in two-flavor $O(4)$-symmetric investigations
\cite{Schaefer:2006ds}, which implicitly assume a maximal
$U(1)_A$-symmetry breaking if the remaining chiral (pseudo)scalar
multiplets, the $\eta$ and $\vec a$ fields, are neglected.
Interestingly, without the determinant, the endpoint is moved in the
opposite direction towards smaller chemical potentials and larger
temperatures if additionally mesonic fluctuations are taken into
account.

\begin{figure*}[t!]
  \centering 
  \subfigure[with $\ua$-symmetry breaking] {\includegraphics[width=6.2cm]{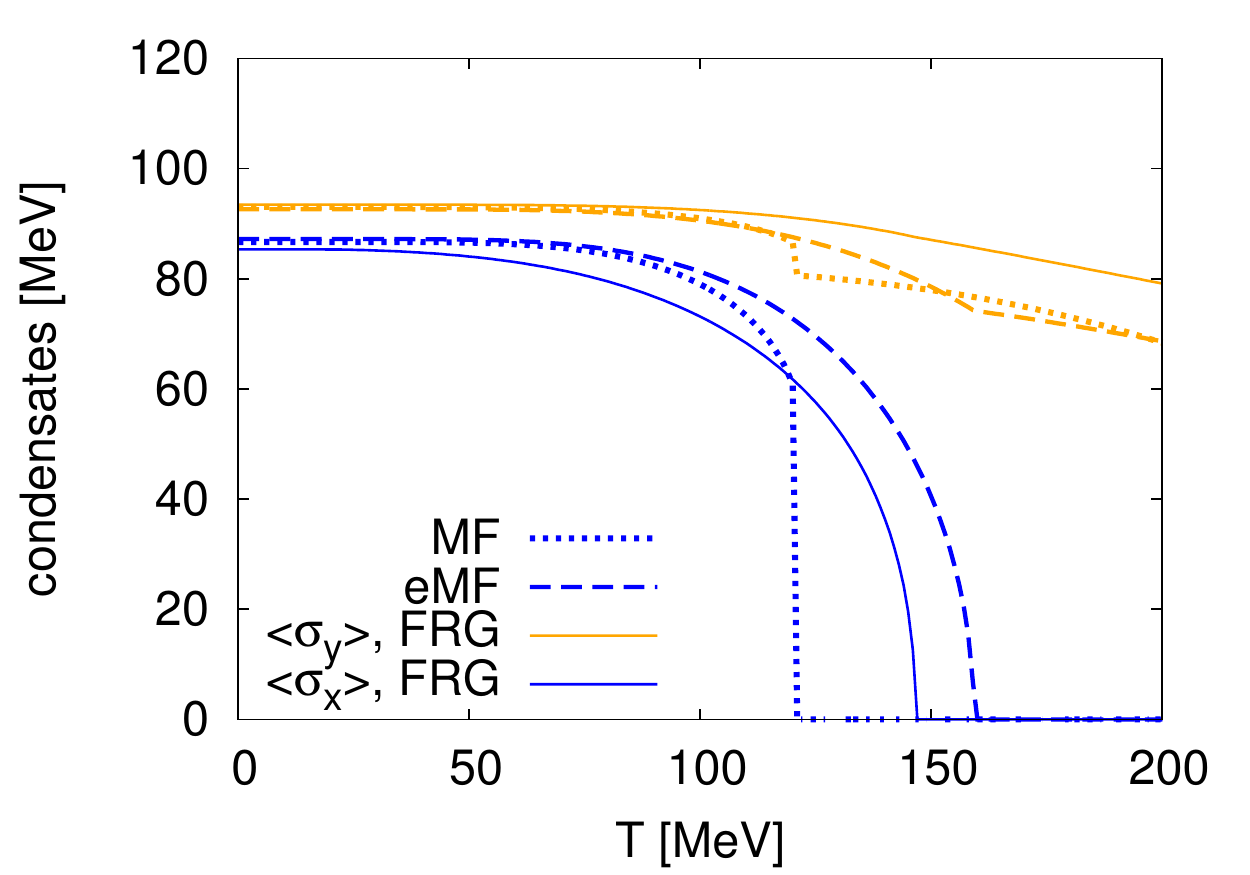}}
  \hfill
  \subfigure[without $\ua$-symmetry breaking] {\includegraphics[width=6.2cm]{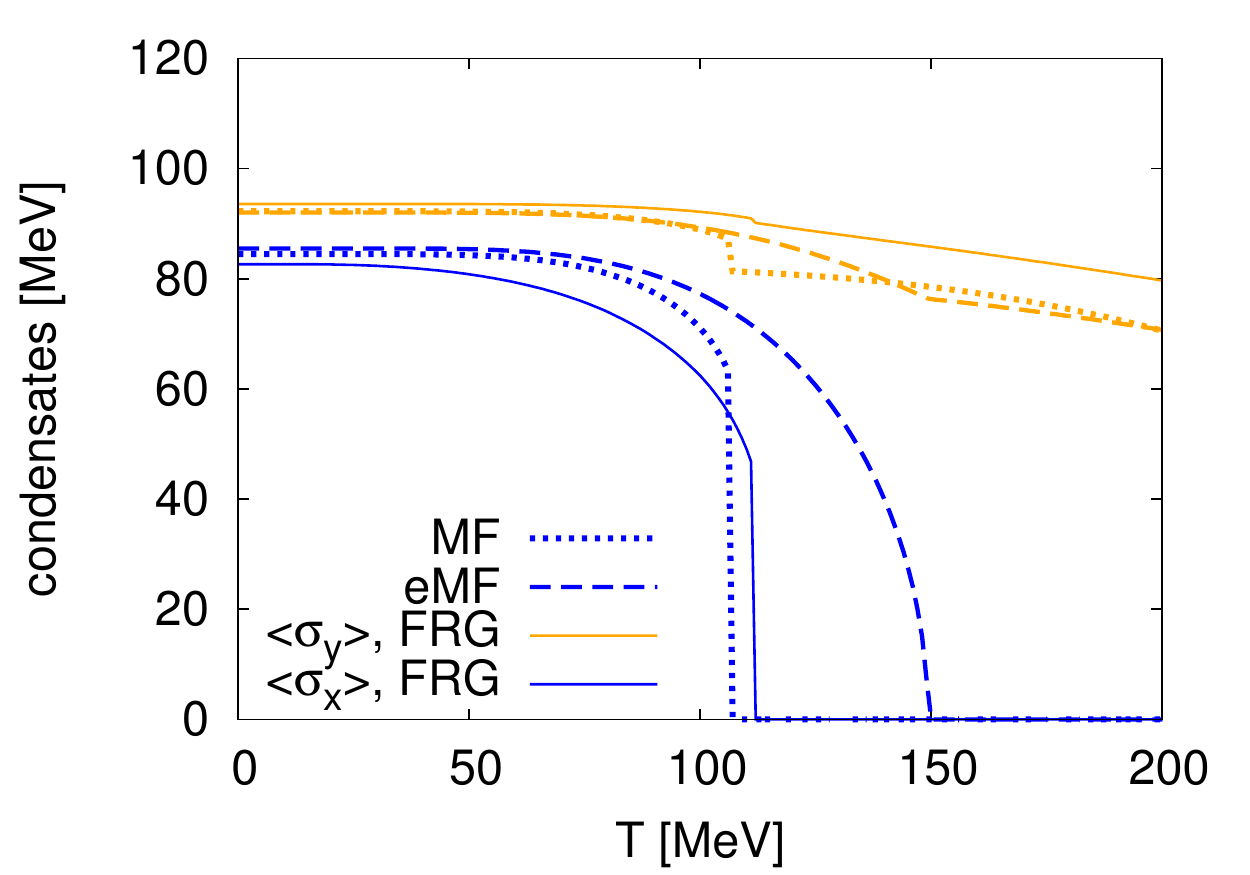}} 
  \caption{\label{fig:frg_mf_condensates_chilim_mu0} Condensates for
    $\mu=0$ similar to Fig.~\ref{fig:frg_condensates_phys_mu0} for the
    light chiral limit $c_x\rightarrow 0$.}
\end{figure*}

It is enlightening to extend the analysis and study the quark mass
sensitivity of the chiral transition with a focus on the role of
fluctuations together with the axial $\ua$-symmetry.  In a RG context
with an $\epsilon$-expansion it has been argued that in a purely
bosonic theory for vanishing anomaly the chiral phase transition
should be of fluctuation-induced first order for $N_f \ge 2$ in the
$SU(3)$-symmetric chiral limit \cite{Pisarski1984a}. Including the
anomaly term the order does not change and the phase transition
remains first order for $N_f \ge 3$. However, the case of $N_f =2$
massless flavors is special. If the temperature-dependence of the
coupling of the, in this case quadratic, Kobayashi-Maskawa-'t~Hooft
determinant is negligible the phase transition can be second order
with $O(4)$ criticality.  In other words a temperature-independent
$\ua$-symmetry breaking can smoo\-th\-en the transition from first to
second order in the two-flavor chiral limit, see also \cite{Grahl:2013pba, Pelissetto:2013hqa}
for recent investigations.  The opposite happens for
three flavors and the first-order transition can become even stronger
in the three-flavor chiral limit.

In Fig.~(\ref{fig:frg_mf_condensates_chilim_mu0}) we show the
non-strange and strange condensates in the light chiral limit,
$c_x\rightarrow 0$ and $c_y \neq 0$ for $\mu=0$ similar to
Fig.~\ref{fig:frg_condensates_phys_mu0}. In agreement with previous
works \cite{Schaefer:2008hk} a first-order transition is found
independent of the $\ua$-symmetry breaking in the standard mean-field
approximation. Going beyond mean-field by including the fermionic
vacuum contribution (eMF) the transition changes to second order.
With the FRG the $\ua$ anomaly has a significant influence: with the
determinant we find a second-order transition and without a
first-order transition.

The results on the quark mass sensitivity of the chiral transition
with and without the $\ua$-symmetry are consistent with the results of
\cite{Pisarski1984a} under the assumption that the
Kobayashi-Maskawa-'t~Hooft determinant behaves qualitatively like a
two-flavor determinant in the case of a physical strange quark
mass. This is also indicated by the fact that the strange condensate
is affected only mildly at the light chiral transition. As a
consequence, the cubic determinant for $N_f=3$ behaves like the
quadratic two-flavor Kobayashi-Maskawa-'t~Hooft term.  Furthermore,
these findings elucidate why the system is ``more critical'' in the
absence of the determinant, i.e., why in this case the critical
endpoint is pushed towards larger temperature and smaller chemical
potential if mesonic fluctuations are taken into account.

Finally, we want to point out that a first order transition emerges in
the chiral limit only if the chiral invariant $\tilde\rho_2$ is taken
into account which is most important for similar two-flavor
investigations.

\section{Summary and conclusions}

The influence of quantum and thermal fluctuations on the chiral
three-flavor phase transition with and without an axial $\ua$-symmetry
breaking is investigated in the framework of the functional
renormalization group with a quark-meson model truncation. Different
mean-field approximations where certain fluctuations are neglected,
are confronted to the full FRG analysis. The $\ua$-symmetry breaking
is effectively implemented by a Kobayashi-Maskawa-'t~Hooft determinant
with a constant coupling strength.

We find a strong dependence of the location of the critical endpoint
on the $\ua$-symmetry breaking if mesonic fluctuations are taken into
account.  With a broken $\ua$-symmetry the endpoint is pushed towards
smaller temperature and larger quark chemical potentials which is in
contrast to corresponding investigations within mean-field
approximations.

In the limit of vanishing light quark masses with a physical strange
quark mass the transition is first-order for a $\ua$-symmetric theory
but second-order with a constant $\ua$-symmetry breaking
Kobayashi-Mas\-ka\-wa-'t~Hooft determinant.  It will be fascinating to
see the results for an improved truncation with a temperature- and
scale-dependent $\ua$-symmetry breaking implementation whose outcomes
should lie in between our two findings.

Both findings, the influence of the determinant on the location of the
critical endpoint as well as the order of the chiral transition, can
be traced back to an effective mass term induced by the $\ua$-symmetry
breaking term in the light flavor sector of the theory.\\[1ex]

\newpage

M.M. acknowledges support by the FWF through DK- W1203-N16, the
Helmholtz Alliance HA216/EMMI, and the BMBF grant
OSPL2VHCTG. B.-J.S. acknowledges support by the FWF grant
P24780-N27. This work is supported by the Helmholtz International
Center for FAIR within the LOEWE program of the State of Hesse.

\bibliography{./rg_references.bib}

\begin{thebibliography}{29}
\expandafter\ifx\csname natexlab\endcsname\relax\def\natexlab#1{#1}\fi
\expandafter\ifx\csname bibnamefont\endcsname\relax
  \def\bibnamefont#1{#1}\fi
\expandafter\ifx\csname bibfnamefont\endcsname\relax
  \def\bibfnamefont#1{#1}\fi
\expandafter\ifx\csname citenamefont\endcsname\relax
  \def\citenamefont#1{#1}\fi
\providecommand{\bibinfo}[2]{#2}
\providecommand{\eprint}[2][]{\url{#2}}

\bibitem{Csorgo2010}
\bibinfo{author}{\bibfnamefont{T.}~\bibnamefont{Csorgo}},
  \bibinfo{author}{\bibfnamefont{R.}~\bibnamefont{Vertesi}}, \bibnamefont{and}
  \bibinfo{author}{\bibfnamefont{J.}~\bibnamefont{Sziklai}},
  \bibinfo{journal}{Phys. Rev. Lett.} \textbf{\bibinfo{volume}{105}},
  \bibinfo{pages}{182301} (\bibinfo{year}{2010}).

\bibitem{Vertesi:2009wf}
\bibinfo{author}{\bibfnamefont{R.}~\bibnamefont{Vertesi}},
  \bibinfo{author}{\bibfnamefont{T.}~\bibnamefont{Csorgo}}, \bibnamefont{and}
  \bibinfo{author}{\bibfnamefont{J.}~\bibnamefont{Sziklai}},
  \bibinfo{journal}{Phys.Rev.} \textbf{\bibinfo{volume}{C83}},
  \bibinfo{pages}{054903} (\bibinfo{year}{2011}).

\bibitem{Bazavov:2012qja}
\bibinfo{author}{\bibfnamefont{A.}~\bibnamefont{Bazavov}} \bibnamefont{et~al.}
  (\bibinfo{collaboration}{HotQCD Collaboration}), \bibinfo{journal}{Phys.Rev.}
  \textbf{\bibinfo{volume}{D86}}, \bibinfo{pages}{094503}
  (\bibinfo{year}{2012}).

\bibitem{Cossu:2012gm}
\bibinfo{author}{\bibfnamefont{G.}~\bibnamefont{Cossu}},
  \bibinfo{author}{\bibfnamefont{S.}~\bibnamefont{Aoki}},
  \bibinfo{author}{\bibfnamefont{S.}~\bibnamefont{Hashimoto}},
  \bibinfo{author}{\bibfnamefont{T.}~\bibnamefont{Kaneko}},
  \bibinfo{author}{\bibfnamefont{H.}~\bibnamefont{Matsufuru}},
  \bibnamefont{et~al.}, \bibinfo{journal}{PoS}
  \textbf{\bibinfo{volume}{LATTICE2011}}, \bibinfo{pages}{188}
  (\bibinfo{year}{2011}).

\bibitem{Cossu:2013uua}
\bibinfo{author}{\bibfnamefont{G.}~\bibnamefont{Cossu}},
  \bibinfo{author}{\bibfnamefont{S.}~\bibnamefont{Aoki}},
  \bibinfo{author}{\bibfnamefont{H.}~\bibnamefont{Fukaya}},
  \bibinfo{author}{\bibfnamefont{S.}~\bibnamefont{Hashimoto}},
  \bibinfo{author}{\bibfnamefont{T.}~\bibnamefont{Kaneko}},
  \bibnamefont{et~al.} (\bibinfo{year}{2013}).

\bibitem{Buchoff:2013nra}
\bibinfo{author}{\bibfnamefont{M.~I.} \bibnamefont{Buchoff}},
  \bibinfo{author}{\bibfnamefont{M.}~\bibnamefont{Cheng}},
  \bibinfo{author}{\bibfnamefont{N.~H.} \bibnamefont{Christ}},
  \bibinfo{author}{\bibfnamefont{H.~T.} \bibnamefont{Ding}},
  \bibinfo{author}{\bibfnamefont{C.}~\bibnamefont{Jung}}, \bibnamefont{et~al.}
  (\bibinfo{year}{2013}).

\bibitem{Cohen:1996ng}
\bibinfo{author}{\bibfnamefont{T.~D.} \bibnamefont{Cohen}},
  \bibinfo{journal}{Phys.Rev.} \textbf{\bibinfo{volume}{D54}},
  \bibinfo{pages}{1867} (\bibinfo{year}{1996}).

\bibitem{Lee:1996zy}
\bibinfo{author}{\bibfnamefont{S.~H.} \bibnamefont{Lee}} \bibnamefont{and}
  \bibinfo{author}{\bibfnamefont{T.}~\bibnamefont{Hatsuda}},
  \bibinfo{journal}{Phys.Rev.} \textbf{\bibinfo{volume}{D54}},
  \bibinfo{pages}{1871} (\bibinfo{year}{1996}).

\bibitem{Birse:1996dx}
\bibinfo{author}{\bibfnamefont{M.~C.} \bibnamefont{Birse}},
  \bibinfo{author}{\bibfnamefont{T.~D.} \bibnamefont{Cohen}}, \bibnamefont{and}
  \bibinfo{author}{\bibfnamefont{J.~A.} \bibnamefont{McGovern}},
  \bibinfo{journal}{Phys.Lett.} \textbf{\bibinfo{volume}{B388}},
  \bibinfo{pages}{137} (\bibinfo{year}{1996}).

\bibitem{Dunne:2010gd}
\bibinfo{author}{\bibfnamefont{G.~V.} \bibnamefont{Dunne}} \bibnamefont{and}
  \bibinfo{author}{\bibfnamefont{A.}~\bibnamefont{Kovner}},
  \bibinfo{journal}{Phys.Rev.} \textbf{\bibinfo{volume}{D82}},
  \bibinfo{pages}{065014} (\bibinfo{year}{2010}).

\bibitem{Meggiolaro:2013swa}
\bibinfo{author}{\bibfnamefont{E.}~\bibnamefont{Meggiolaro}} \bibnamefont{and}
  \bibinfo{author}{\bibfnamefont{A.}~\bibnamefont{Morda}},
  \bibinfo{journal}{Phys. Rev. D} \textbf{\bibinfo{volume}{88}},
  \bibinfo{pages}{096010} (\bibinfo{year}{2013}).

\bibitem{Pisarski1984a}
\bibinfo{author}{\bibfnamefont{R.~D.} \bibnamefont{Pisarski}} \bibnamefont{and}
  \bibinfo{author}{\bibfnamefont{F.}~\bibnamefont{Wilczek}},
  \bibinfo{journal}{Phys. Rev.} \textbf{\bibinfo{volume}{D29}},
  \bibinfo{pages}{338} (\bibinfo{year}{1984}).

\bibitem{Grahl:2013pba}
\bibinfo{author}{\bibfnamefont{M.}~\bibnamefont{Grahl}} \bibnamefont{and}
  \bibinfo{author}{\bibfnamefont{D.~H.} \bibnamefont{Rischke}},
  \bibinfo{journal}{Phys.Rev.} \textbf{\bibinfo{volume}{D88}},
  \bibinfo{pages}{056014} (\bibinfo{year}{2013}).

\bibitem{Pelissetto:2013hqa}
\bibinfo{author}{\bibfnamefont{A.}~\bibnamefont{Pelissetto}} \bibnamefont{and}
  \bibinfo{author}{\bibfnamefont{E.}~\bibnamefont{Vicari}},
  \bibinfo{journal}{Phys.Rev.} \textbf{\bibinfo{volume}{D88}},
  \bibinfo{pages}{105018} (\bibinfo{year}{2013}).

\bibitem{Kobayashi:1970ji}
\bibinfo{author}{\bibfnamefont{M.}~\bibnamefont{Kobayashi}} \bibnamefont{and}
  \bibinfo{author}{\bibfnamefont{T.}~\bibnamefont{Maskawa}},
  \bibinfo{journal}{Prog.Theor.Phys.} \textbf{\bibinfo{volume}{44}},
  \bibinfo{pages}{1422} (\bibinfo{year}{1970}).

\bibitem{'tHooft:1976fv}
\bibinfo{author}{\bibfnamefont{G.}~\bibnamefont{'t~Hooft}},
  \bibinfo{journal}{Phys.Rev.} \textbf{\bibinfo{volume}{D14}},
  \bibinfo{pages}{3432} (\bibinfo{year}{1976}).

\bibitem{Schaefer:2008hk}
\bibinfo{author}{\bibfnamefont{B.-J.} \bibnamefont{Schaefer}} \bibnamefont{and}
  \bibinfo{author}{\bibfnamefont{M.}~\bibnamefont{Wagner}},
  \bibinfo{journal}{Phys. Rev.} \textbf{\bibinfo{volume}{D79}},
  \bibinfo{pages}{014018} (\bibinfo{year}{2009}).

\bibitem{Wetterich:1992yh}
\bibinfo{author}{\bibfnamefont{C.}~\bibnamefont{Wetterich}},
  \bibinfo{journal}{Phys. Lett.} \textbf{\bibinfo{volume}{B301}},
  \bibinfo{pages}{90} (\bibinfo{year}{1993}).

\bibitem{Gies:2001nw}
\bibinfo{author}{\bibfnamefont{H.}~\bibnamefont{Gies}} \bibnamefont{and}
  \bibinfo{author}{\bibfnamefont{C.}~\bibnamefont{Wetterich}},
  \bibinfo{journal}{Phys.Rev.} \textbf{\bibinfo{volume}{D65}},
  \bibinfo{pages}{065001} (\bibinfo{year}{2002}).

\bibitem{Gies:2002hq}
\bibinfo{author}{\bibfnamefont{H.}~\bibnamefont{Gies}} \bibnamefont{and}
  \bibinfo{author}{\bibfnamefont{C.}~\bibnamefont{Wetterich}},
  \bibinfo{journal}{Phys.Rev.} \textbf{\bibinfo{volume}{D69}},
  \bibinfo{pages}{025001} (\bibinfo{year}{2004}).

\bibitem{Floerchinger:2009uf}
\bibinfo{author}{\bibfnamefont{S.}~\bibnamefont{Floerchinger}}
  \bibnamefont{and}
  \bibinfo{author}{\bibfnamefont{C.}~\bibnamefont{Wetterich}},
  \bibinfo{journal}{Phys.Lett.} \textbf{\bibinfo{volume}{B680}},
  \bibinfo{pages}{371} (\bibinfo{year}{2009}).

\bibitem{Pawlowski:2010ht}
\bibinfo{author}{\bibfnamefont{J.~M.} \bibnamefont{Pawlowski}},
  \bibinfo{journal}{AIP Conf.Proc.} \textbf{\bibinfo{volume}{1343}},
  \bibinfo{pages}{75} (\bibinfo{year}{2011}).

\bibitem{Braun:2011pp}
\bibinfo{author}{\bibfnamefont{J.}~\bibnamefont{Braun}},
  \bibinfo{journal}{J.Phys.} \textbf{\bibinfo{volume}{G39}},
  \bibinfo{pages}{033001} (\bibinfo{year}{2012}).

\bibitem{Litim:2001up}
\bibinfo{author}{\bibfnamefont{D.~F.} \bibnamefont{Litim}},
  \bibinfo{journal}{Phys. Rev.} \textbf{\bibinfo{volume}{D64}},
  \bibinfo{pages}{105007} (\bibinfo{year}{2001}).

\bibitem{Mitter:2013fxa}
\bibinfo{author}{\bibfnamefont{M.}~\bibnamefont{Mitter}} \bibnamefont{and}
  \bibinfo{author}{\bibfnamefont{B.-J.} \bibnamefont{Schaefer}},
  \href{http://arxiv.org/abs/1308.3176}{{\tt arXiv:1308.3176 [hep-ph]}}.

\bibitem{Schaefer:2006sr}
\bibinfo{author}{\bibfnamefont{B.-J.} \bibnamefont{Schaefer}} \bibnamefont{and}
  \bibinfo{author}{\bibfnamefont{J.}~\bibnamefont{Wambach}},
  \bibinfo{journal}{Phys. Part. Nucl.} \textbf{\bibinfo{volume}{39}},
  \bibinfo{pages}{1025} (\bibinfo{year}{2008}).

\bibitem{mmanomalie}
\bibinfo{author}{\bibfnamefont{M.}~\bibnamefont{Mitter}},
  \bibinfo{author}{\bibfnamefont{B.-J.} \bibnamefont{Schaefer}},
  \bibinfo{author}{\bibfnamefont{N.}~\bibnamefont{Strodthoff}},
  \bibnamefont{and} \bibinfo{author}{\bibfnamefont{L.}~\bibnamefont{von
  Smekal}}, \bibinfo{journal}{in prep.}  (\bibinfo{year}{2014}).

\bibitem{Pawlowski:1996ch}
\bibinfo{author}{\bibfnamefont{J.~M.} \bibnamefont{Pawlowski}},
  \bibinfo{journal}{Phys. Rev.} \textbf{\bibinfo{volume}{D58}},
  \bibinfo{pages}{045011} (\bibinfo{year}{1998}).

\bibitem{Schaefer:2006ds}
\bibinfo{author}{\bibfnamefont{B.-J.} \bibnamefont{Schaefer}} \bibnamefont{and}
  \bibinfo{author}{\bibfnamefont{J.}~\bibnamefont{Wambach}},
  \bibinfo{journal}{Phys.Rev.} \textbf{\bibinfo{volume}{D75}},
  \bibinfo{pages}{085015} (\bibinfo{year}{2007}).

\end{thebibliography}

\end{document}